\documentclass[pre,twocolumn,amsmath,amssymb,notitlepage]{revtex4}
\pdfoutput=1
\usepackage{dcolumn}
\usepackage{lipsum}
\usepackage{graphicx}
\usepackage{hyperref}
\usepackage{float}
\usepackage{bm}
\usepackage[T1]{fontenc}
\usepackage{color}
\usepackage{url}
\usepackage[bf]{subfigure}
\usepackage{rotating}

\begin{document}

\title{Modular Dynamics of Financial Market Networks}

\author{Filipi N. Silva$^1$}
\email{filipinascimento@gmail.com}
\author{Cesar H. Comin$^1$}
\email{chcomin@gmail.com}
\author{Thomas K. DM. Peron$^1$}
\author{Francisco A. Rodrigues$^2$}
\author{Cheng Ye$^3$}
\author{Richard C. Wilson$^3$}
\author{Edwin Hancock$^3$}
\author{Luciano da F. Costa$^1$}
\email{ldfcosta@gmail.com}
\affiliation{$^1$Instituto de F\'{\i}sica de S\~{a}o Carlos, Universidade de S\~{a}o Paulo, S\~{a}o Carlos, S\~ao Paulo, Brazil\\
$^2$Departamento de Matem\'{a}tica Aplicada e Estat\'{i}stica, Instituto de Ci\^{e}ncias Matem\'{a}ticas e de Computa\c{c}\~{a}o,Universidade de S\~{a}o Paulo, Caixa Postal 668,13560-970 S\~{a}o Carlos, S\~ao Paulo, Brazil\\
$^3$Department of Computer Science, University of York, York, YO10 5GH, United Kingdom}


\begin{abstract}
The financial market is a complex dynamical system composed of a large variety of intricate relationships between several entities, such as banks, corporations and institutions. At the heart of the system lies the stock exchange mechanism, which establishes a time-evolving network of trades among companies and individuals. Such network can be inferred through correlations between time series of companies stock prices, allowing the overall system to be characterized by techniques borrowed from network science. Here we study the presence of communities in the inferred stock market network, and show that the knowledge about the communities alone can provide a nearly complete representation of the system topology. This is done by defining a simple null model, a randomized version of the  studied network sharing only the sizes and interconnectivity between communities observed. We show that many topological characteristics of the inferred networks are carried over the networks generated by the null model. In particular, we find that in periods of instability, such as during a financial crisis, the network strays away from a state of well-defined community structure to a much more uniform topological organization. We show that the framework presented here provides a good null model representation of topological variations taking place in the market during crises. Also, the general approach used in this work can be extended to other systems.
\end{abstract}

\maketitle

\section{Introduction}
\label{sec:introduction}

The quantitative analysis of real complex systems is mostly focused on the study of time series, since the full knowledge of the underlying equations of a given dynamic system is hard or usually impossible to obtain from static data~\cite{abarbanel1996analysis,diks1999nonlinear,kantz2004nonlinear,sprott2003chaos}. This situation is frequently found in the analysis of a number of applicable domains, including financial markets~\cite{mantegna2000introduction}, physiological data~\cite{kantz2011nonlinear}, climate modeling~\cite{schreiber1999interdisciplinary} and several other systems~\cite{schreiber1999interdisciplinary}. 

Many tools based on nonlinear dynamics (e.g., fractal dimensions, Lyapunov exponents and recurrence properties) have been developed to analyze time series originating from complex systems~\cite{schreiber1999interdisciplinary}. However, besides such conventionally studied approaches, recently, concepts of network science have been applied to this problem too. Generally, most available methods map time series into the network domain, so that a set of network measurements can be used to calculate the statistical properties of the system. These methods can be quite different and include cycle networks~\cite{zhang2006complex}, transition networks~\cite{nicolis2005dynamical}, $k$-nearest-neighbors~\cite{shimada2008analysis}, visibility-graphs~\cite{lacasa2008time} and also recurrence networks~\cite{marwan2009complex,donner2010recurrence} which is a natural extension of the traditional recurrence analysis of dynamical systems~\cite{eckmann1987recurrence,marwan2007recurrence}. Ref.~\cite{donner2010recurrence} provides a brief review of the analysis of time series using complex networks techniques, also comparing the approaches and pointing to possible pitfalls and limitations of their applications.

The aforementioned methods analyze a single time series using complex networks. However, one could also tackle complex systems consisting of sets of time series, rather than just a single sample. Examples of such systems include the stock market in which each asset has a time evolving market price~\cite{mantegna1999hierarchical,tumminello2010correlation,tumminello2007spanning,bonanno2004networks,ye2015thermodynamic,onnela2003dynamics}; temporal spatio-grid climate data, where each grid point has a time series associated to a given climate variable (e.g., temperature, wind velocity, etc.)~\cite{donges2009complex,donges2009backbone,yamasaki2008climate,gozolchiani2008pattern}; and also grid temporal data collected from electroencephalograms (EEG) or functional magnetic resonance images (fMRI)~\cite{bullmore2009complex,fallani2011multiple}. Each of these systems are represented by discrete elements whose physical interactions can be inferred from the corresponding time series. 

Given their discrete nature, the motivation for using the network approach to analyze such systems is natural. Network techniques usually establish the connections (interactions) by quantifying the statistical similarities between each discrete element, yielding a so-called functional network~\cite{bullmore2009complex}. Moreover, the complex network approach to complex systems also provides an effective and comprehensive set of visualization tools, which allows important insights on the relation between the network structure and the underlying system behavior. The benefits of network visualization can be realized, for instance, in the study of biological~\cite{tong2004global,goh2007human,costanzo2010genetic,guimera2005functional}, social~\cite{wasserman1994social,newman2010networks} and transportation networks~\cite{barthelemy2011spatial}, among others~\cite{newman2010networks,boccaletti2006complex}. 

Naturally, these functional networks can evolve with time, and techniques can be developed for network construction in order to analyze the underlying complex system. In such an approach, the main objective is the detection of extreme events that significantly modify the network structure, which can in turn be related to critical events in the underlying complex system. For instance, in the temporal analysis of climate networks one important issue is detecting extreme events caused by El Ni\~no-Southern Oscillation (ENSO)~\cite{clarke2008introduction,power2013robust,yamasaki2008climate,gozolchiani2008pattern,yamasaki2009climate,ludescher2013improved,ludescher2014very,radebach2013disentangling} or the occurrences of Monsoons~\cite{malik2010spatial,malik2012analysis}. In financial market networks, the effect of financial instabilities in the cluster organization of stocks is of interest~\cite{mantegna1999hierarchical,tumminello2010correlation,tumminello2007spanning,bonanno2004networks,onnela2003dynamics,bordino2012web,peron2011collective,battiston2013complex,piccardi2011clustering,fenn2012dynamical,fenn2009dynamic}. In each case, the occurrence of extreme events is inferred from the detection of anomalies in the time series originating from the network evolution.

\begin{figure}[!tbp]
  \begin{center}
  \includegraphics[width=1.0\linewidth]{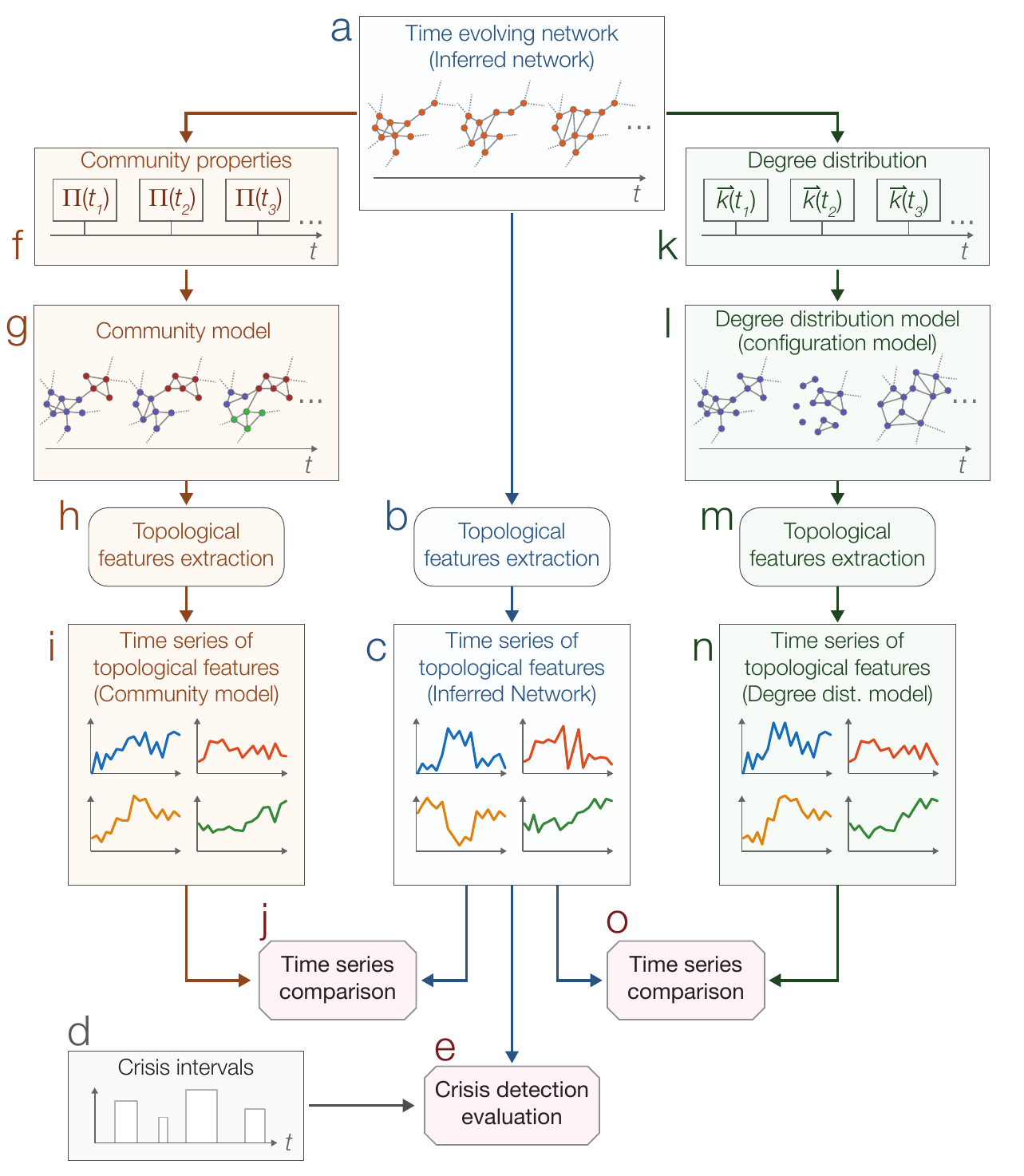} 
  \caption{(Color online) Schematic representation of the different steps of our methodology. Please refer to the text for an explanation about each step.}
  ~\label{f:methods}
  \end{center}
\end{figure}

In this paper, we compare the set of time series obtained from the evolution of the \emph{inferred network} with those obtained from a null model of the same network, in which only information about its communities is used. By \emph{null model} we understand a family of randomized networks reproducing a subset of features of the original networks~\cite{newman2006modularity}. The methodology outlined in our paper is described in detail in Fig.~\ref{f:methods}. Using a financial market data set, we obtain the a time-evolving network (step a), which is inferred from the Pearson correlation coefficient between company daily closing stock values. A set of time series of topological measurements is obtained from the evolving network (step b), and used to characterize the evolution of network structure (step c). Using the widely documented crisis periods (step d), we evaluate the extent of which topological features correctly reflects the changes in network structure during different crises (step e). 

In order to create a community null model of the evolution of the stock market network, a set of community properties is obtained (step f). At each time step, a random network with same average degree and community structure as the corresponding inferred network is created using a stochastic blockmodel~\cite{newman2011stochastic}, generating a time-evolving community network (step g). Finally, once the time series of topological measurements are extracted (steps h and i), the network evolution derived from the random network with communities can be compared with the time series of topological measurements of the inferred network (step j). Such comparison needs a proper scale to allow us to conclude how similar the network generated from the community model is to the inferred network. This is done by using a well-known random network model, known as configuration model~\cite{newman2010networks}, to also represent the inferred network. That is, the degree distribution of the network at each day is obtained (step k), and used to generate daily networks having the same degree distribution as the inferred ones (step l). Topological features of these networks are obtained (steps m and n) and compared with the same features obtained for the inferred networks (step o). If the community null model can provide a more suitable representation of the inferred networks, we expect the comparison done at step j to display a closer agreement between the time series than the comparison done at step o.

The suggested analyses for the stock market networks are built upon the stochastic blockmodel and are based solely on the community structure of the time evolving inferred network. As far as we know, our approach is different from other approaches based on network community, as most of them are focused on testing community detection algorithms~\cite{girvan2002community,lancichinetti2008benchmark,sawardecker2009detection}. While more sophisticated community models, such as the degree-corrected blockmodel~\cite{newman2011stochastic}, could also have been used, they would incorporate not only the community structure but also other characteristics of the original system, such as degree distribution and average path length. Therefore it could be difficult to disentangle the effects of these properties on the results.
It is important to highlight that the purpose of the current work is not to predict the behavior of the financial market network, but to understand how informative is the community structure when traditional topological features are used to analyze such networks. This is performed by comparing a set of properties between the inferred networks and the respective networks generated by the null model. Such null model can then be used to isolate the influence of community structure changes on other topological properties, such as assortativity and transitivity. Therefore, it provides a precise means to verify which properties are a trivial consequence of variations in the community structure.

We validate our framework by analyzing functional networks constructed through statistical similarities between stocks traded at the New York Stock Exchange (NYSE). We show that the financial crashes are characterized by the presence of well-defined changes to the community structure, whereas outside these critical periods the network topology is composed by communities that remain stable for long periods. Accordingly, we found that the observed values of many topological properties of the financial market network are a direct consequence of the community structure of the system.

This paper is organized as follows. In Sec. II we specify how the time evolving network of the financial market is inferred, and describe some basic community changes observed in the network during a single crisis period. In Sec. III we present the methodology used to recover the market topology using daily mixing matrices. We highlight the relevance of the community structure for the financial market characterization. Finally, in Sec. IV we present the conclusions of the study.

\section{The Time-evolving Stock Market Network}
\label{sec:realnetworks}

The stock market database consists of daily prices of 3799 stocks traded on the New York Stock Exchange. The stocks prices were obtained from the Yahoo! financial database (http://finance.yahoo.com). We selected $N=348$ stocks from this set, which are the stocks for which there are historical data from January 1986 to February 2011. For these stocks, we obtained $C_p=6008$ closures prices per stock over the trading period.

In the financial market networks analyzed in this paper, the nodes correspond to stocks and the links quantify the statistical similarity between the time series associated to the stock closure prices evolution. In particular, in order to quantify the similarity between two time series, we adopt the Pearson correlation coefficient
\begin{equation}
\rho_{ij}=\frac{\left\langle Y_{i}Y_{j}\right\rangle -\left\langle Y_{i}\right\rangle \left\langle Y_{j}\right\rangle }{\sqrt{\left(\left\langle Y_{i}^{2}\right\rangle -\left\langle Y_{i}\right\rangle ^{2}\right)\left(\left\langle Y_{j}^{2}\right\rangle -\left\langle Y_{j}\right\rangle ^{2}\right)}}, 
\label{Eq:correlation_coefficient_rho}
\end{equation} 
where $Y_i(t)$ is the logarithm of return, i.e., $Y_i(t) = \ln P_i(t) - \ln P_i(t-1)$ is the price return, where $P_i(t)$ is the closing price of the $i$-th stock at day $t$. The advantage
of using $Y(t)$ is that nonlinear and stochastic transformations are not needed in order to correct for some common trends in the data \cite{mantegna1999introduction}. In addition, $Y(t)$ is independent of inflation or discount factors. By calculating Eq.~\ref{Eq:correlation_coefficient_rho} between all pairs of stocks we obtain a fully connected weighted graph in which the link weights are given by $\rho_{ij}$. However, in order to analyze the network structure according to the stronger connections, we discard links whose weights are below a threshold $\epsilon$. This leads to a network defined by the adjacency matrix $A_{ij} = \Theta(\rho_{ij} - \epsilon) - \delta_{ij}$, where $\Theta(\cdot)$ is the Heaviside function and $\delta_{ij}$ the Kronecker delta. 

\begin{figure*}[!htbp]
  \begin{center}
  \includegraphics[width=0.9\linewidth]{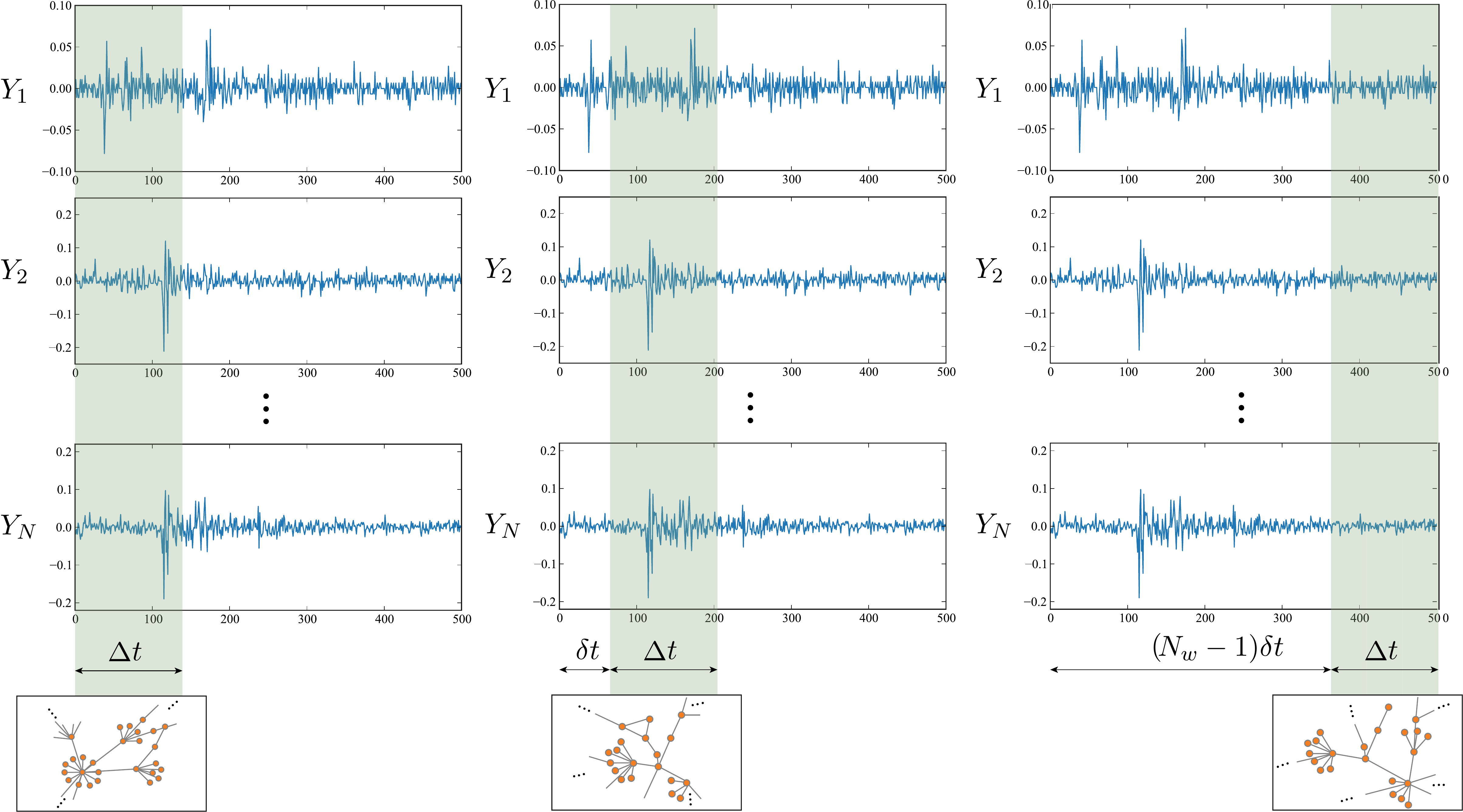} 
  \caption{(Color online) Diagram illustrating the method to construct the financial market networks. The network is constructed by calculating the correlations between the stocks returns $Y_i$ ($i=1,2,...,N$) inside a time window of length $\Delta t$. Next, by shifting this time window by amounts $\delta t$ until the end of the database is reached, we obtain the network evolution.}
  ~\label{f:network_construction}
  \end{center}
\end{figure*}

The analysis of the time evolution of the inferred financial market network is illustrated in Fig.~\ref{f:network_construction}. First we set a time window of length $\Delta t = 30$ days inside which the correlations between stocks are calculated creating the network. Next, by sequentially sliding the window by $\delta t=1$ we are able to generate a sequence of networks corresponding to the evolution of the market. More precisely, the first network $\mathcal{N}^{(1)}$ is constructed by calculating the correlations of the time series between $t^{(1)}_1=1$ and $t^{(1)}_2=31$, the second network $\mathcal{N}^{(2)}$ is constructed considering data between instants $t^{(2)}_1=2$ and $t^{(2)}_2=32$, and the $n$-th network encompasses the prices between $t^{(n)}_1 = 1 + (n-1)\delta t$ and $t^{(n)}_2 = t^{(n)}_1 + \Delta t$. Therefore, $n$ represents the index of the analyzed business day and takes value in the range $\left[1,N_w\right]$, where $N_w=C_p-\Delta t$. The threshold $\epsilon_t$ used to define which edges are kept changes for each day. The value of $\epsilon_t$ is set so that $f=10\%$ of the possible $N(N-1)/2$ edges are kept at each day. That is, the average degree of the network is fixed. In Fig. S1 of the supplementary material we show the evolution of modularity for different choices of $f$. It is clear that there are no significant changes in the results if $f$ belongs to the range $[5\%,20\%]$.

\begin{figure*}[!htbp]
  \begin{center}
  \includegraphics[width=0.9\linewidth]{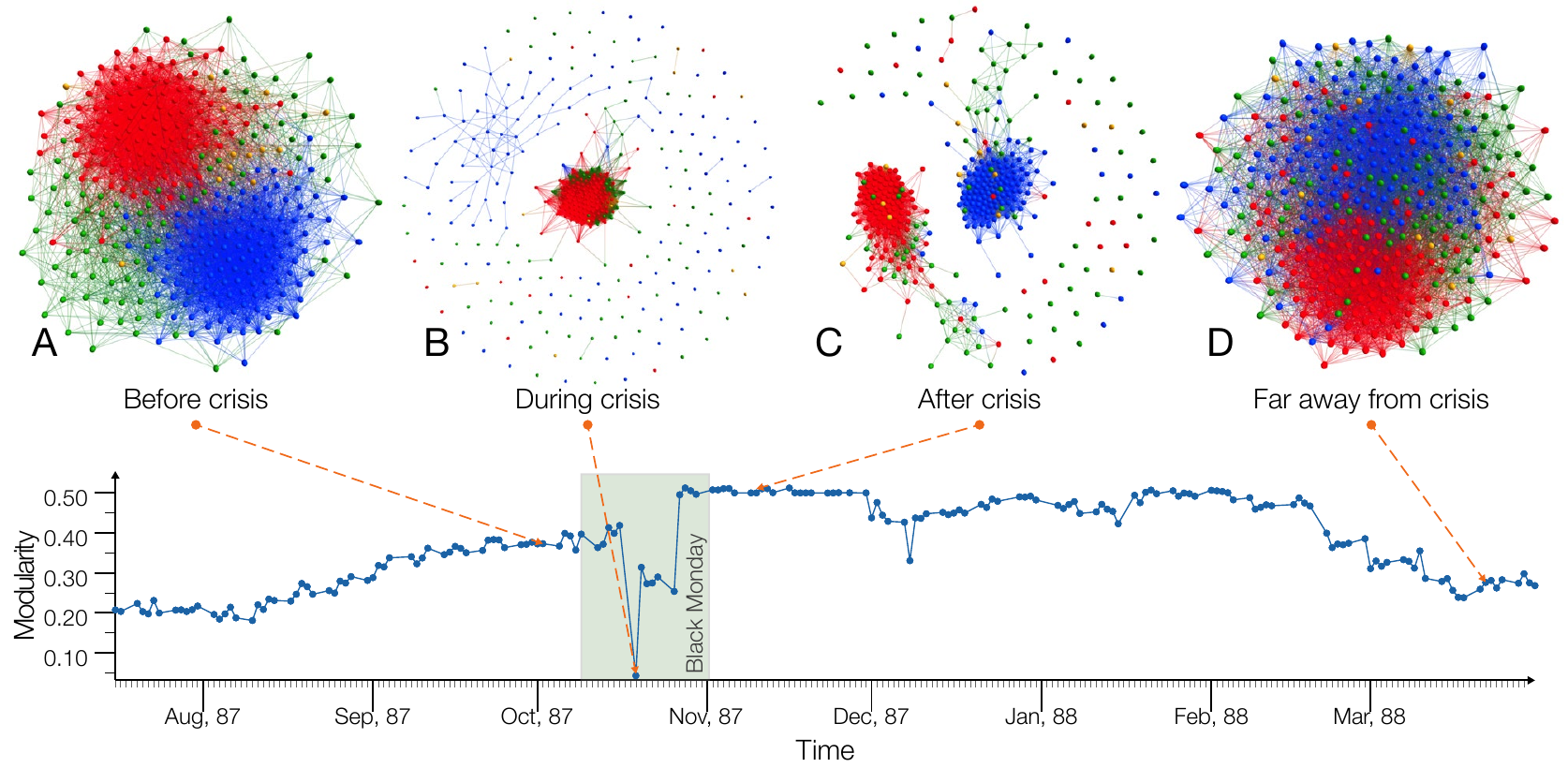} 
  \caption{(Color online) Modularity of the NYSE networks in distinct times during and around the Black Monday crisis~\cite{onnela2003dynamic,onnela2003dynamics}. We also show a visualization of the network at four specific days. Node colors correspond to the community structure found for network \textbf{A}. We note that the average degrees of networks \textbf{A}, \textbf{B}, \textbf{C} and \textbf{D} are the same.}
  ~\label{f:crisis_communities_visu}
  \end{center}
\end{figure*}

In order to better understand the evolution of the financial market network, it is useful to first visualize how its structure is organized near critical points. A useful and clearly defined event that can be used as a reference for the effect of financial instabilities in the network structure is the famous Black Monday crisis, which occurred in October 19, 1987~\cite{onnela2003dynamic,onnela2003dynamics}. After employing the methodology described above for network construction, the community structure of the network at each business day was found using the multilevel community detection algorithm~\cite{blondel2008fast}. The resulting modularity of the detected communities is shown in Fig.~\ref{f:crisis_communities_visu}. We note that distinct connected components of the network are always assigned to different communities. In Fig.~\ref{f:crisis_communities_visu}, we also show network visualizations corresponding to four different instants of time, where each node color represents a different community. Furthermore, in order to correctly observe the communities evolution, the community membership of each node is calculated only in network \textbf{A} of Fig.~\ref{f:crisis_communities_visu}, so that the community assignment is kept fixed for the visualization of networks \textbf{B}, \textbf{C} and \textbf{D}.

From the figure, it is clear that before the crisis the modular structure is mainly composed of two predominant communities. As the network approaches the crisis the network changes drastically, and the community structure substantially vanishes. Only a highly connected cluster at the center of the network remains. Note that since we fix the average degree, the number of edges in networks \textbf{A} and \textbf{B} is the same, only the connectivity is changed. At this epoch, most stocks are disconnected, meaning that the prices evolve without strong correlations. During the crisis, the remaining connected components exhibit a more homogeneous structure, as suggested by the lower values of modularity. Similar repercussions of the Black Monday were reported in the temporal analysis of the Minimum Spanning Tree (MST) constructed from the NYSE data~\cite{onnela2003dynamic,onnela2003dynamics}. The authors observed that the asset tree shrinks, i.e. the normalized tree length decreases, reducing the topological distances between the stocks. This result also agrees with other findings on the structural organization of financial market networks~\cite{peron2012thestructure,yan2014stability,kumar2012correlation,peron2011collective,fenn2011temporal,mcdonald2008impact}.  It is also interesting to note in Fig.~\ref{f:crisis_communities_visu} that, throughout the considered period, the communities preserve most of their membership composition, even in very long periods after the crash, when the network becomes reconnected again.

\begin{figure*}[!htbp]
  \begin{center}
  \includegraphics[]{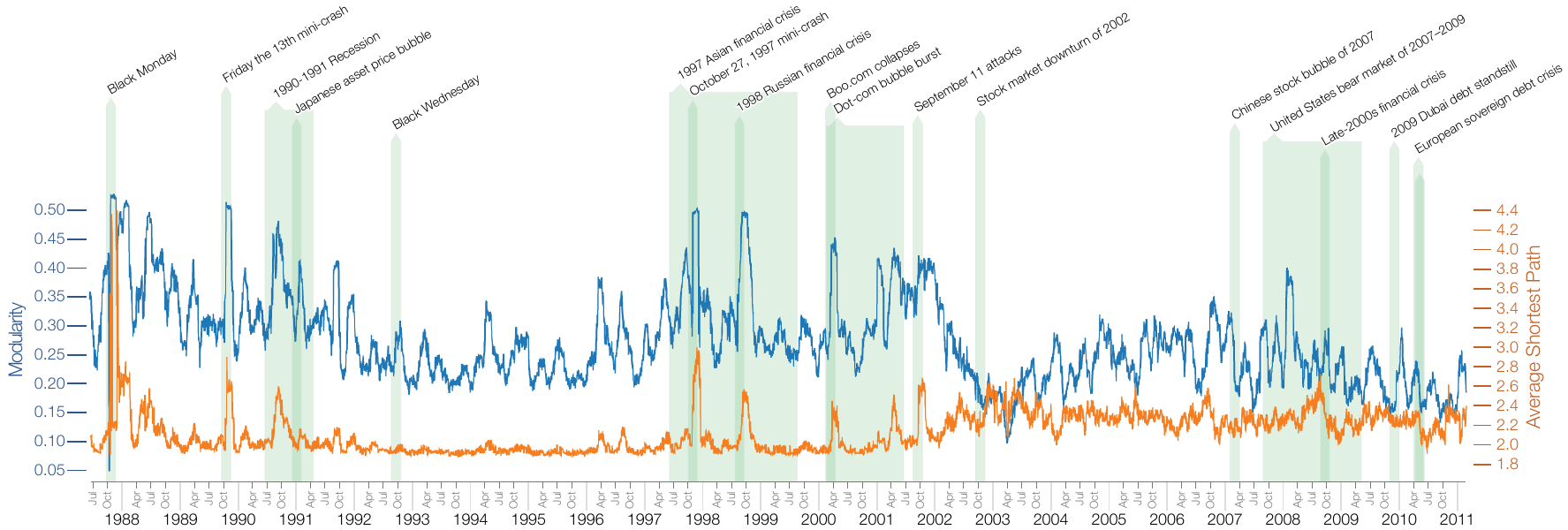}
\begin{minipage}{\linewidth}
  \caption[von neumann]{(Color online) Modularity (blue) and average shortest path length (orange) of the NYSE networks for all businnes days in our dataset. Known periods of crisis and bear market trends\footnote{\scriptsize \url{http://www.wsj.com/articles/SB119239926667758592}\\ \url{http://www.telegraph.co.uk/finance/2773265/Billionaire-who-broke-the-Bank-of-England.html} \\ \url{http://www.guardian.co.uk/technology/2005/may/16/media.business} \\ \url{http://news.bbc.co.uk/onthisday/hi/dates/stories/april/26/newsid_2503000/2503271.stm} \\ \url{http://www.theguardian.com/world/2003/apr/09/russia.artsandhumanities} \\ Retrieved 09 May, 2015 } \cite{Guhathakurta2015Ec} are shown. We considered only crises that had influence on the US stock market. Note that the dates specified by the source are approximate and, in general, the duration of a crisis is not well-defined in the literature, with the exception of marked crises such as the \emph{Black Monday}.}  \label{f:crisis_mod_aspl}
\end{minipage}
  \end{center}
\end{figure*}

In order to quantitatively investigate the relationship between a financial crisis and network community structure, we analyze a set of well known crisis periods. These periods are marked alongside the curve of the modularity in Fig.~\ref{f:crisis_mod_aspl}, for all business days of our dataset. We also show in the same plot the evolution of the average shortest path length of the network. This measurement seems to be less sensitive to fluctuations outside crisis periods, and therefore is able to provide additional confirmation that there are significant structural changes during crises. For each considered crisis, we observe an increase of modularity around the time span of the crisis. This indicates that indeed the modularity can capture topological network properties related to the financial crisis occurring in the system. For instance, the $1990\rightarrow1991$ recession seems to be closely related to the period of high values of modularity between mid 1990 and early 1991. 

\begin{figure*}[!htbp]
  \begin{center}
  \includegraphics[width=18.5cm]{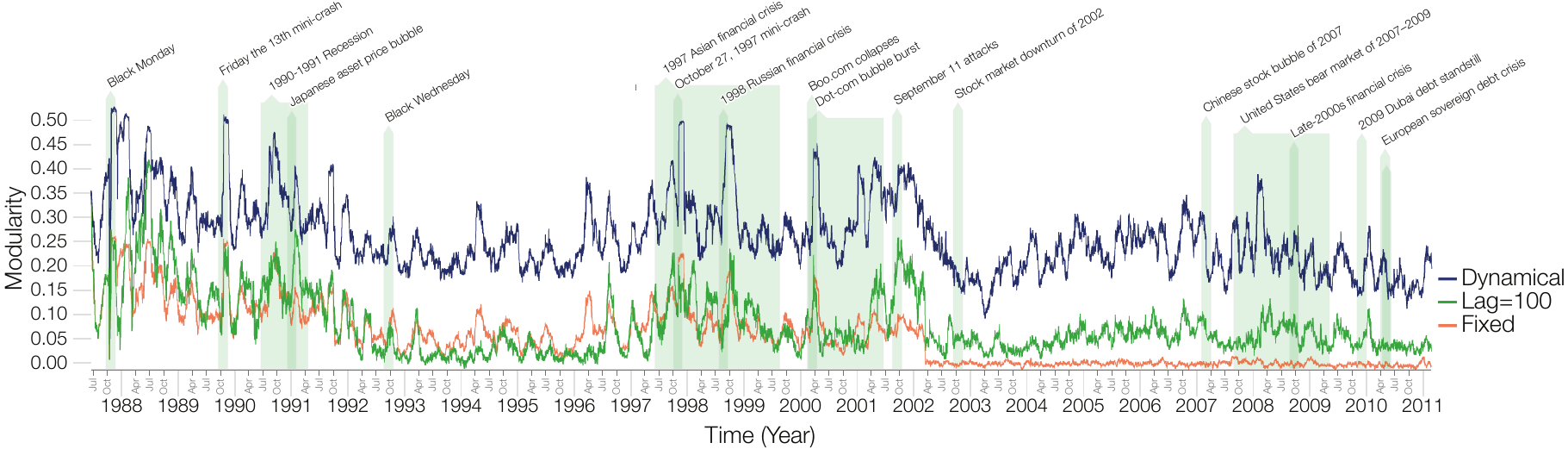} 
  \caption{(Color online) Evolution of the dynamical modularity, fixed modularity and lagged modularity of the stock market network with time. Crisis periods are shown in a manner similar to Fig.~\ref{f:crisis_mod_aspl}.}
  ~\label{f:modularityLagged100}
  \end{center}
\end{figure*}

An important aspect of analyzing the modularity of time-evolving networks is that changes in modularity can be caused by two distinct factors. It may be due to changes in the community membership of the nodes or because of changes in the network topology. In order to study the relative impact of both of these aspects, we define three types of modularity, namely \emph{dynamical}, \emph{fixed} and \emph{lagged}. For the \emph{dynamical modularity}, at each time step node memberships are recalculated using the multilevel community detection algorithm \cite{blondel2008fast}. Therefore the modularity reflects the best network partition found by the algorithm at each day. In Fig.~\ref{f:modularityLagged100} we show in blue the evolution of the dynamical modularity for the entire time series. We see that this measurement contains spikes, that is, increased values of dynamical modularity during short periods, while there is no long-time variation of the modularity. Although the dynamical modularity is a useful indicator of the overall association between stock prices, it provides no information about the actual changes occurring inside the communities. For example, the membership of the nodes can change along time steps without modifying the dynamical modularity.

Therefore, we also characterize the community evolution of the financial market by using a \emph{fixed modularity} defined as follows. The community structure found on the first time step defines the membership of the nodes for all time steps, and the modularity is calculated based on this membership. In Fig.~\ref{f:modularityLagged100} we show in orange the evolution of the fixed modularity. It is likely that, over a long time, stocks may evolve into different communities, so fixed modularity tends to decrease with time. It is clear that spikes are less pronounced than when using the dynamical modularity. Also, there is a clear transition of the fixed modularity around the year 2002, which means that there was a noticeable change of the financial market at this epoch. We will discuss this transition on the following section. Nevertheless, an obvious problem with the fixed modularity is that there is nothing to distinguish the starting point of the time series, where the community composition and node membership was calculated. This problem can be solved by defining a \emph{lagged modularity}, in which the memberships are calculated at time $t-t_{\Delta}$ and used to determine the modularity at time $t$. If $t_{\Delta}=0$, we simply recover the dynamical modularity. For large $t_{\Delta}$ the values tend to the fixed modularity, since when $t-t_{\Delta}<0$ we consider the memberships calculated in the first time step. To some extent, the value $t_{\Delta}$ is related to the memory of the market, that is, the extent to which its structure changed during the interval $t_{\Delta}$. In Fig.~\ref{f:modularityLagged100} we show the lagged modularity for $t_{\Delta}=100$ business days. In Fig. S2 of the supplementary material we present the lagged modularity for different values of $t_{\Delta}$. In a manner similar to the fixed modularity, the spikes of the lagged modularity are also less pronounced than when using the dynamical modularity. This means that although the market contains pronounced communities during the spike periods, the composition of the communities are varying with time.

\section{Topology Characterization from Daily Mixing Matrices}
\label{sec:networkModel}

In the stock exchange system, companies are typically organized in several sectors (i.e. industry, services, banking, etc.) which may possess very distinctive relationship patterns and time series behavior. The inferred networks are also expected to exhibit a modular organization based on node sectors. In the previous section, we showed qualitatively that the community structure on the inferred networks may be related to periods of crisis. 

To further investigate the importance of the community structure in this system, we use a stochastic blockmodel to generate networks that preserve only information about the community structure of the inferred networks. Starting from the set of inferred networks obtained from the NYSE dataset, for each time step $t$ we obtain the community structure $C(t)$ by employing the multilevel community detection method proposed in \cite{blondel2008fast}. From the resulting partitioning of the nodes, we obtain a set of community mixing matrices $\Pi(t)$. The elements in $\Pi_{\alpha \beta}(t)$ correspond to the number of connections between each pair of communities $(\alpha,\beta)$ at time $t$, where $\alpha \in C(t)$ and $\beta \in C(t)$. Additionally, the number of nodes within each community $\alpha$, $|\alpha|$, is also considered. We define our null model in terms of a sequence of stochastic mixing matrices, $\pi(t)$, that can be expressed by a normalization of $\Pi_{\alpha\beta}(t)$:
\begin{equation}
\pi_{\alpha \beta}(t) = \begin{cases}
	\frac{2 \Pi_{\alpha \beta}(t)}{|\alpha| (|\alpha|-1) }& \text{if } \alpha = \beta \\[.5em]
	\frac{\Pi_{\alpha \beta}(t)}{|\alpha| |\beta|} & \text{if } \alpha \neq \beta 
	 \end{cases}.
\label{eq:normalized_mixing}
\end{equation}
The parameter $\pi(t)$ corresponds to the same probability of connection between groups on the simplest blockmodel~\cite{newman2011stochastic}. However, here we took a different approach to the generative model, we apply a wiring process more similar to the way that simple Erd\H{o}s R\'enyi networks are constructed. For each instant, a network is initially generated as a fully disconnected graph with $\sum_{\alpha\,\in\,C(t)} |\alpha|$ nodes. Each node is assigned to a community $\alpha \in C(t)$ in agreement with the given community sizes, $|\alpha|$. Next, pairs of nodes $(i,j)$, with community memberships $i \in \alpha$ and $i \in \beta$, are connected according to the probability $\pi_{\alpha \beta}(t)$ . The resulting networks contain communities formed by uniform random graphs (similar to the Erd\H{o}s R\'enyi model), and the only characteristic they share \emph{a priori} with the inferred networks is the community structure. Since we recalculate the communities at each time step, the topology is based on the daily mixing matrices of the inferred networks.

The remainder of this section is devoted to comparing the topological properties obtained from our community null model with those obtained for the inferred networks. In addition, we also compare our results with those obtained from networks generated using the configuration model~\cite{newman2010networks}. This model is widely used in the literature to generate a null model of the network under study. This is so because the generated networks are guaranteed to have the same degree distribution as the original ones, while any other structural property not related to individual node degrees is the same as in an uniformly random network. Our aim is to uncover in which conditions the community null model might be a better descriptor of the inferred network topology.

\begin{figure*}[!htbp]
  \begin{center}
  \includegraphics[width=18.1cm]{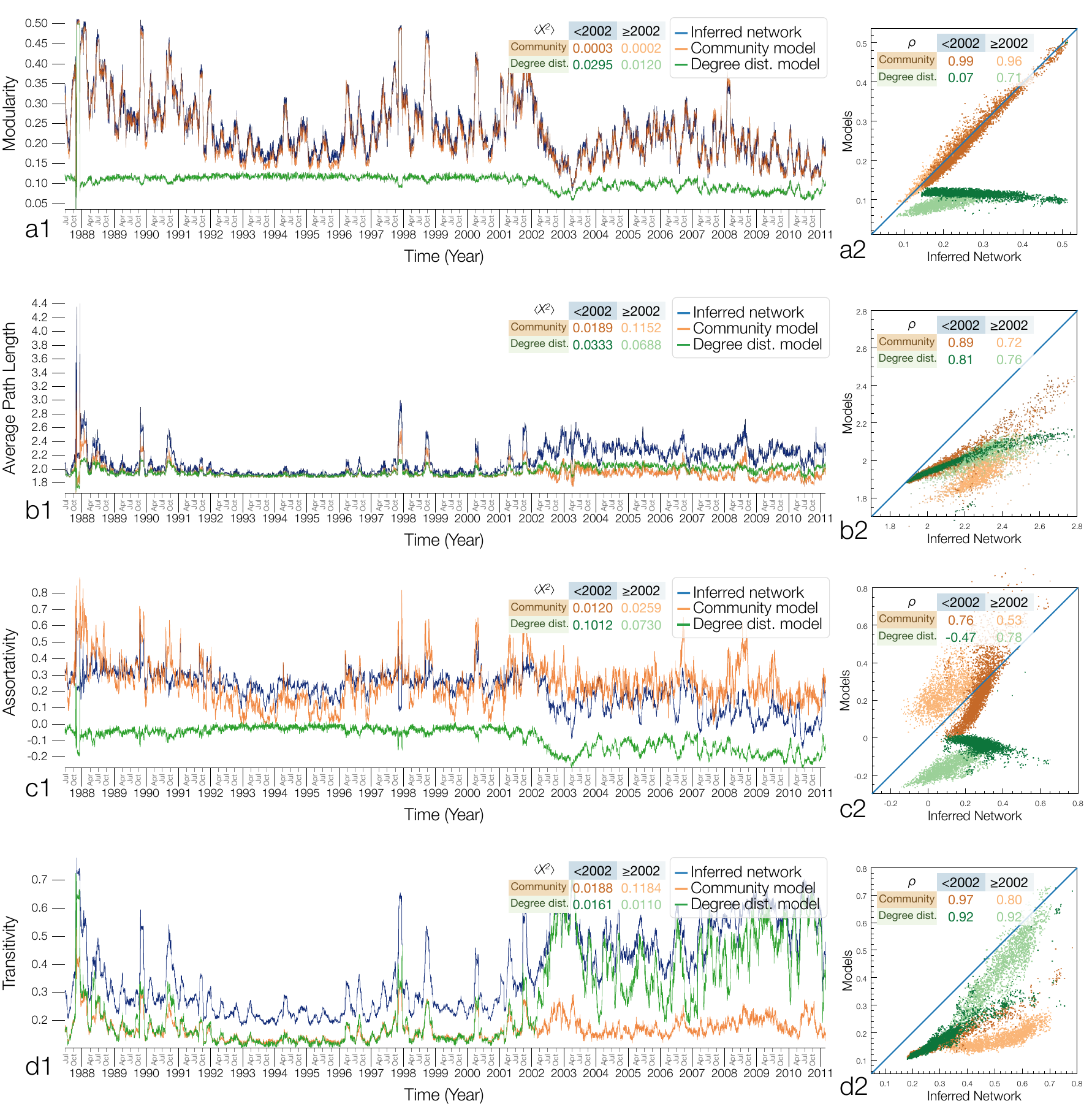} 
  \caption{(Color online) Dynamical modularity, average path length, degree assortativity and transitivity obtained for the NYSE networks (blue), configuration null model (green) and community null model (orange). Plots on the left show the time evolution of the measurements. Plots on the right show the scatter plot between each measurement obtained from the inferred networks and the artificially generated networks. The blue line indicates a $y=x$ relationship. The average squared differences and Pearson correlation coefficients between the inferred and simulated time series are shown as tables included inside each plot. Both measurements are calculated separately for the $p_{1987\rightarrow2002}$ and $p_{2002\rightarrow2011}$ periods.}
  ~\label{f:plotsModelVsRealMeasurementsA}
  \end{center}
\end{figure*}

We measured eight topological characteristics~\cite{costa2007characterization}, namely dynamical modularity~\cite{newman2006modularity}, average shortest path length~\cite{watts1998collective}, degree assortativity~\cite{newman2002assortative}, transitivity~\cite{newman2001scientific}, average betweenness \cite{brandes2001afaster}, clique number~\cite{bomze1999maximum}, righ-club coefficient~\cite{zhou2004rich,colizza2006detecting} and average matching index~\cite{costa2007characterization}. The results for the first four mentioned measurements are shown in Fig.~\ref{f:plotsModelVsRealMeasurementsA}. For brevity, the plots comparing the remaining network characteristics are included in Fig. S3 of the supplementary material.

The dynamical modularity, shown in Fig.~\ref{f:plotsModelVsRealMeasurementsA}(a1), is measured to provide a confirmation that our null model is indeed capturing the community structure of the inferred networks. As expected, the modularity of the networks generated by only fixing the degree distribution, shown in green, is usually low. The only period where such networks display noticeably large values of modularity is during the Black Monday crisis, which happens because the network becomes highly disconnected in this particular crisis. In order to provide a quantitative comparison of the time series, we measure the average squared difference between the calculated values for the inferred network and the community and degree distribution null models. The result is shown as a table in the upper right corner of the plot. In Fig.~\ref{f:plotsModelVsRealMeasurementsA}(a2) we show a scatter plot between the modularity observed in the inferred network and the modularity of the respective networks generated from the community and configuration models. The blue line indicates a $y=x$ relationship, which would be the optimal agreement between the inferred and artificial networks. As noted before, there is an apparent transition of market behavior around the year 2002. Therefore, we also divide the whole series into two intervals, the period $1987\rightarrow2002$, represented by darker markers, and the period $2002\rightarrow2011$, represented by lighter marks. In the same figure, we also show the Pearson correlation coefficient between the measurements for both periods. It is useful to measure both the average squared difference and the Pearson coefficient because while the former considers only the absolute difference between the measurements, the latter indicates how well the two properties vary together. This is useful to account for cases where the two properties display a large difference in absolute value, but their changes along time are still related. In such cases, the null model will still reflect sudden changes in the inferred network topology during crises.

In Fig.~\ref{f:plotsModelVsRealMeasurementsA}(b1) we show the average shortest path length of the inferred and generated networks. This measurement does not change significantly with time, except at epochs where the modularity increases, since a more modular network tends to give on average larger shortest distances. It is also clear that there is a larger deviation from the null model after the year 2002, although the variations between different periods are still represented by the generated networks, as indicated by the large Pearson correlation coefficients indicated in Fig.~\ref{f:plotsModelVsRealMeasurementsA}(b2). 

The degree assortativity of the networks is shown in Fig.~\ref{f:plotsModelVsRealMeasurementsA}(c1), where again there was a significant relationship between the community null model and the inferred network. The assortativity of a uniformly random network should be close to zero, since there is no preferential connectivity between nodes of similar degree. Therefore, one mighty conclude that our null model should produce non-assortative networks, but the community description of a network is so effective that it also constrains the degree assortativity. For example, if one of the network communities is denser than the remaining communities, then the nodes belonging to this community will have large degrees and have a tendency to be more strongly interconnected. This increases the assortativity of the network. The scatter plot comparing the degree assortativity between the inferred and modeled networks (Fig.~\ref{f:plotsModelVsRealMeasurementsA}(c2)) confirm this idea. The community null model follows closely the $y=x$ line, specially in the $1987\rightarrow2002$ period.

The transitivity (shown in Fig.~\ref{f:plotsModelVsRealMeasurementsA}(d1)) is the characteristic that displays the largest absolute deviation between the inferred and artificial community networks, when compared to the respective fixed degree distribution networks. This occurs because the transitivity of uniformly random networks tends to zero for large network sizes~\cite{watts1998collective}. Since the inferred networks actually have a non-zero transitivity, the null model will usually have a lower transitivity values than for the inferred case. Nevertheless, the transitivity variations between time steps are still carried over the simulated networks, since the community structure captured by the null model also has a strong influence on the transitivity. This is confirmed by the large Pearson correlation coefficients obtained for the scatter plot shown in Fig.~\ref{f:plotsModelVsRealMeasurementsA}(d2). It is interesting that the degree distribution alone can provide a good representation of the inferred networks transitivity in the $2002\rightarrow2011$ period, while the community model presents a strong deviation in the transitivity values.

\begin{figure}[!htbp]
  \begin{center}
  \includegraphics[width=0.9\linewidth]{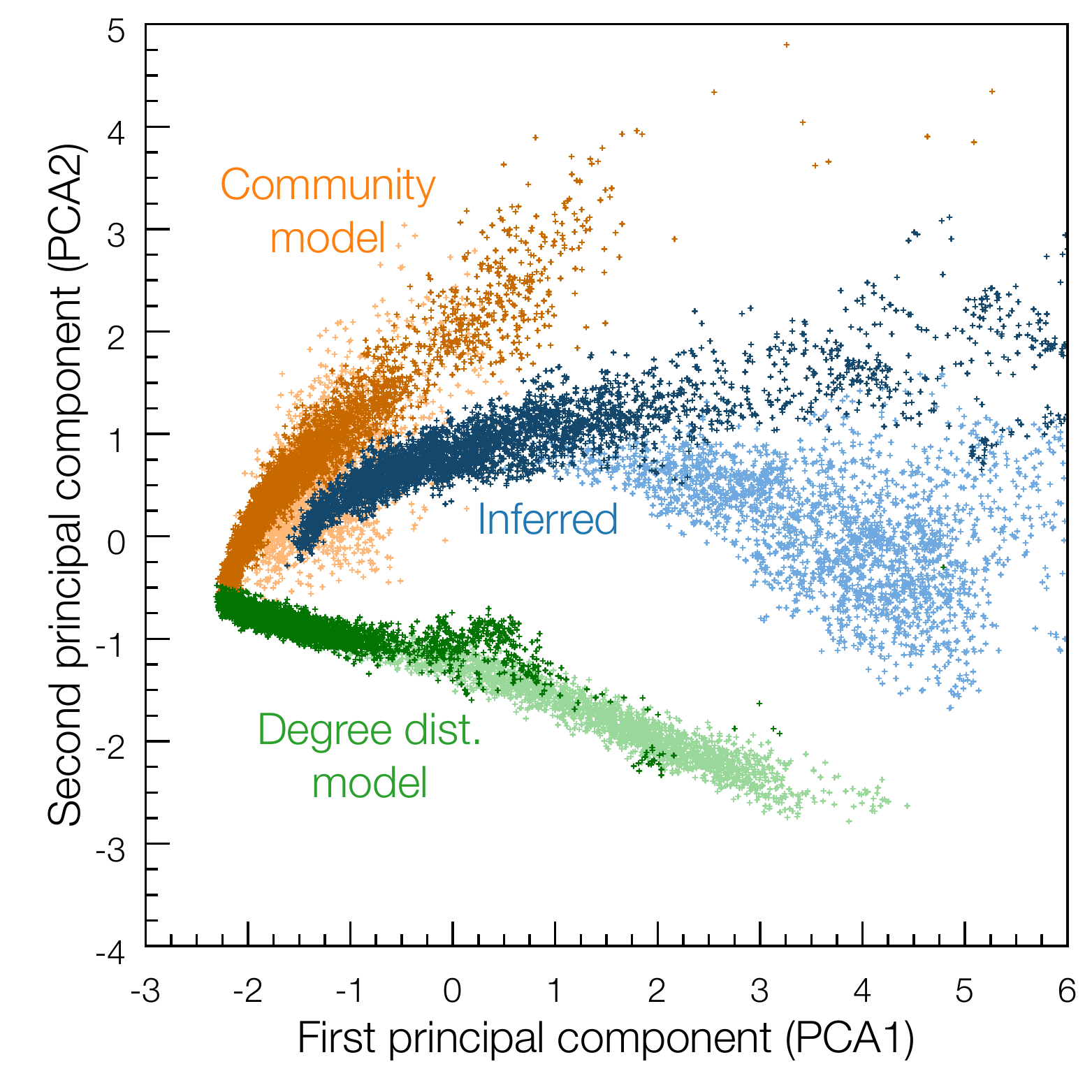} 
  \caption{(Color online) PCA obtained from all considered measurements for the inferred networks (blue), the community preserving null model (orange) and by fixing the degree distribution (green).}
  ~\label{f:pca}
  \end{center}
\end{figure}

In order to summarize the results obtained for the eight measurements calculated over the inferred and artificially generated time-evolving networks, we apply the Principal Component Analysis technique~\cite{jolliffe2002Principal} to the data. The original 8-dimensional space spanned by the obtained network measurements is projected onto a 2-dimensional one, defined by the first two PCA components. The result is shown in Fig.~\ref{f:pca}. We observe that the first principal component (PCA1) is strongly related to changes in the network structure along time, since the main axis of dispersion in the original 8-dimensional space is a consequence of the network evolution. Therefore, the models should be compared according to the second principal component (PCA2). Thus, it is clear that the community null model carries over many characteristics of the inferred networks. Keeping only the daily degree distribution of the inferred networks seems to be an \emph{inefficient} method for reproducing the original network topology. By inefficient we mean that storing the entire daily degree distribution of the network requires much more information than storing the daily mixing matrix describing the community structure.

\begin{figure*}[!htbp]
  \begin{center}
  \includegraphics[width=0.9\linewidth]{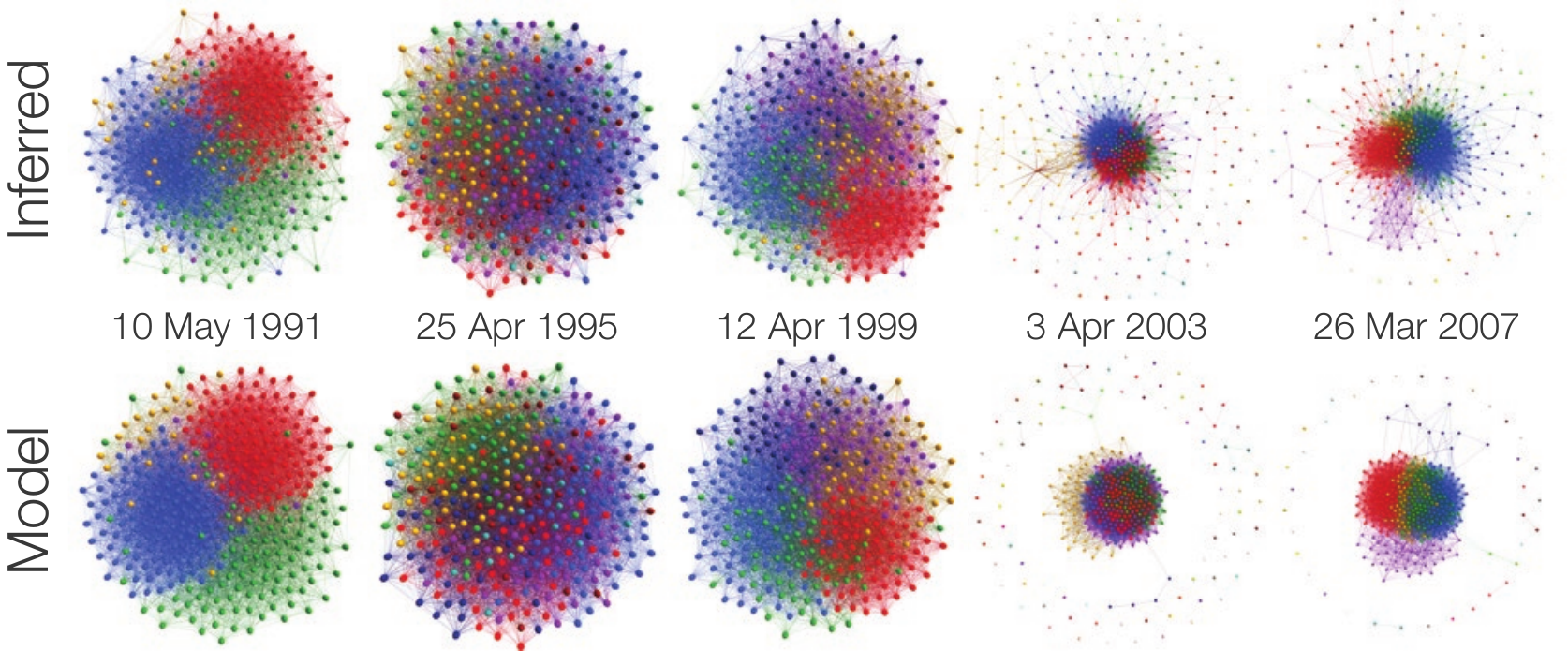} 
  \caption{(Color online) Network visualizations comparing the inferred networks and those generated by the community null model. We sampled the networks every $1000$ time steps. Colors indicate the community membership of nodes.}
  ~\label{f:visualizationComparison}
  \end{center}
\end{figure*}

As a final example, in Fig.~\ref{f:visualizationComparison} we visually compare the inferred networks with those generated by our null model. The inferred networks were sampled every $1000$ time steps between 10 May 1991 and 26 March 2007. To construct the visualization we use a force-directed method based on the Fruchterman-Reingold algorithm~\cite{fruchterman1991graph} for which only the network topology is taken into account, hence no information about the communities is considered aside from nodes colors. Again, we observe that the networks constructed from the mixing matrices are able to qualitatively describe the topology of the inferred financial networks.

\section{Conclusions}
\label{sec:conclusions}

The study of the evolving financial market network is of great interest for many reasons. Besides improving decisions related to industrial development and overall national growth, a reliable description of the evolution of a financial market can provide indicators of an imminent widespread stock value decline, which we refer to as a stock market crisis. In order to infer the underlying financial market network for a group of companies, one widespread technique is to obtain the correlation networks of the stocks exchanged between them~\cite{tumminello2010correlation}. Such inferred networks tend to convey many relevant properties of the actual financial network established by the financial market. In addition, since companies can always be divided into distinct subgroups, related to their respective sector of activity or partnerships, the analysis of the community structure of the financial market is of considerable importance to understanding its dynamics.

The goal of this paper was to show that the community structure of the financial market contains an almost complete description of the dynamics of the system, in the sense that many of its characteristics can be recovered by knowing only the communities over a given time interval. We here demonstrated that by using the mixing matrices obtained from the inferred networks, many network measurements can indeed be recovered by generating a random network following the connectivity pattern indicated by these matrices. Interestingly, after the year 2002 the inferred networks become less related to the community preserving artificial networks generated by the null model. This effect remains the topic of future work. We found no clear explanation for the mechanisms involved in the formation and evolution of the communities.

The results indicate that, on many occasions, it is more suitable to use the community preserving null model to attest the statistical significance of experimental observations on financial market networks. Furthermore, the methodology can be applied to any system where the underlying network can be inferred through similarity measurements. If the system displays a natural representation based on its constituent communities, we expect the framework proposed here to provide an accurate description of topological changes occurring in the system. It also remains to be studied if different pruning techniques~\cite{serrano2009extracting,radicchi2011information}, used to transform a correlation matrix into the adjacency matrix, can improve the results or provide new insights about the data. In addition, other datasets related to the financial  market, such as inter-bank ownership, could provide additional developments about the relevance of community formation during pronounced market crises.


\section*{Acknowledgements}
C. H. Comin thanks FAPESP (Grant No. 11/22639-8) for financial support. F. N. Silva acknowledges CAPES. L. da F. Costa thanks CNPq (Grant no. 307333/2013-2) and NAP-PRP-USP for support. T. K. D. M. Peron acknowledges FAPESP (Grant no. 2012/22160-7) for support. F. A. Rodrigues acknowledges CNPq (grant 305940/2010-4), FAPESP (grant 2011/50761-2 and 2013/26416-9) and NAP eScience - PRP - USP for financial support. This work has been supported also by FAPESP grants 12/50986-7 and 11/50761-2. E. R. Hancock acknowledges Royal Society Wolfson Research Merit Award for support.

\bibliographystyle{apsrev}
\bibliography{paper}

\newcommand{\beginsupplement}{%
        \setcounter{table}{0}
        \renewcommand{\thetable}{S\arabic{table}}%
        \setcounter{figure}{0}
        \renewcommand{\thefigure}{S\arabic{figure}}%
     }
\clearpage
\onecolumngrid

\beginsupplement

\appendix
\section{Supplementary Material of  "Modular Dynamics of Financial Market Networks"}

\begin{figure*}[!htbp]
  \begin{center}
  \includegraphics[width=18.1cm]{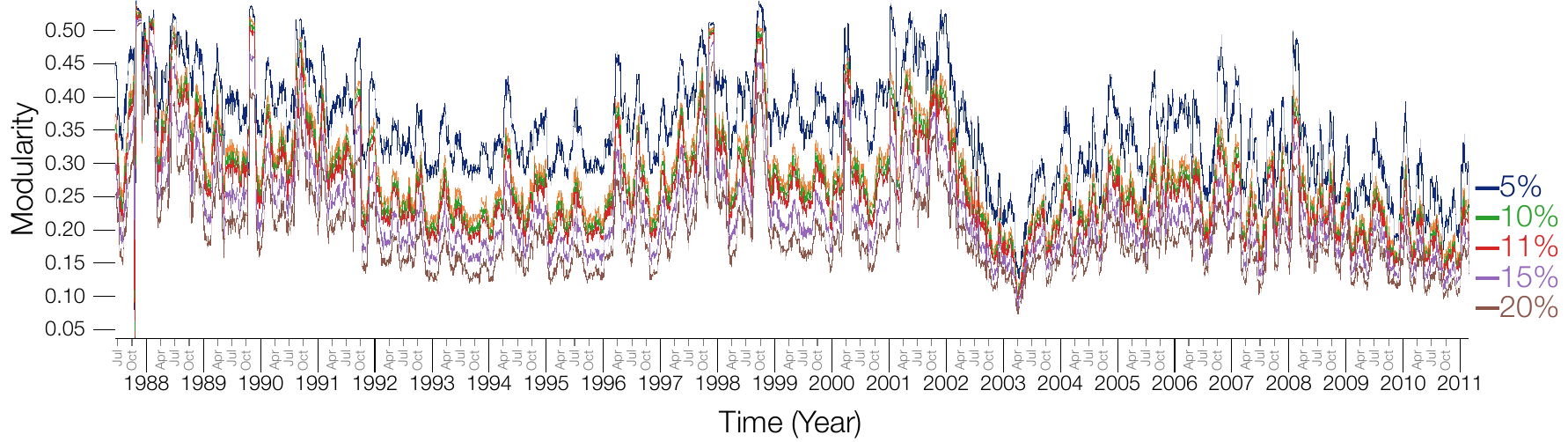} 
  \caption{Effect of the desired percentage of edges $f$ on the dynamical modularity of the inferred network.}
  ~\label{fig:modularityThreshold}
  \end{center}
\end{figure*}

\begin{figure*}[!htbp]
  \begin{center}
  \includegraphics[width=18.5cm]{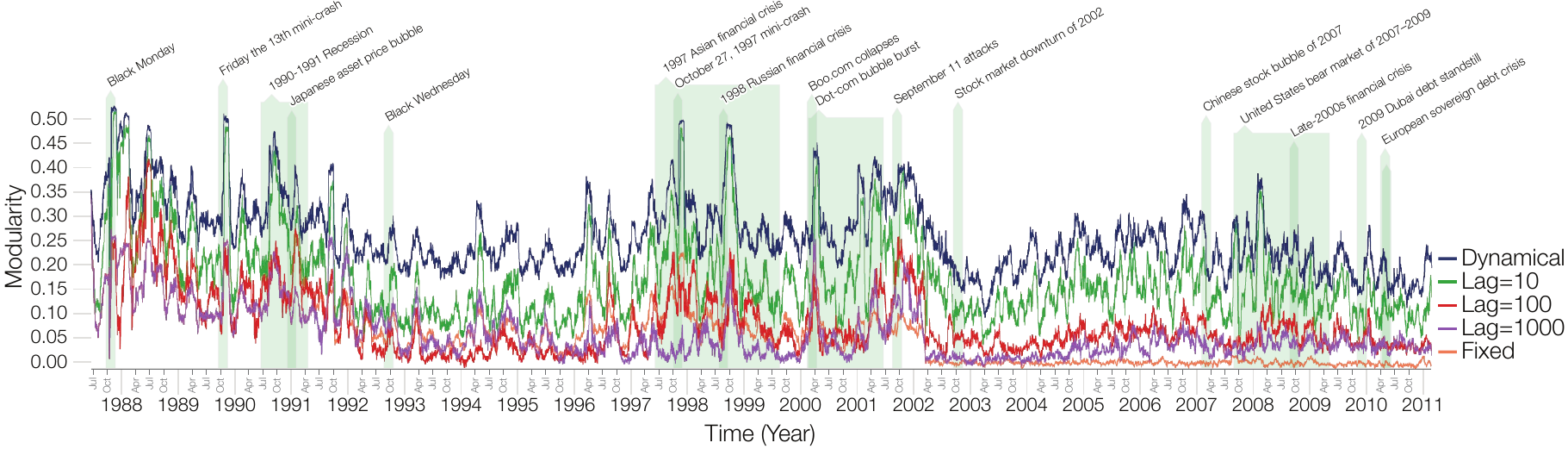} 
  \caption{Evolution of the dynamical, fixed and lagged modularity (considering $3$ lag values) of the stock market network with time.}
  ~\label{f:modularityLaggedAll}
  \end{center}
\end{figure*}

\begin{figure*}[!htbp]
  \begin{center}
  \includegraphics[width=18.1cm]{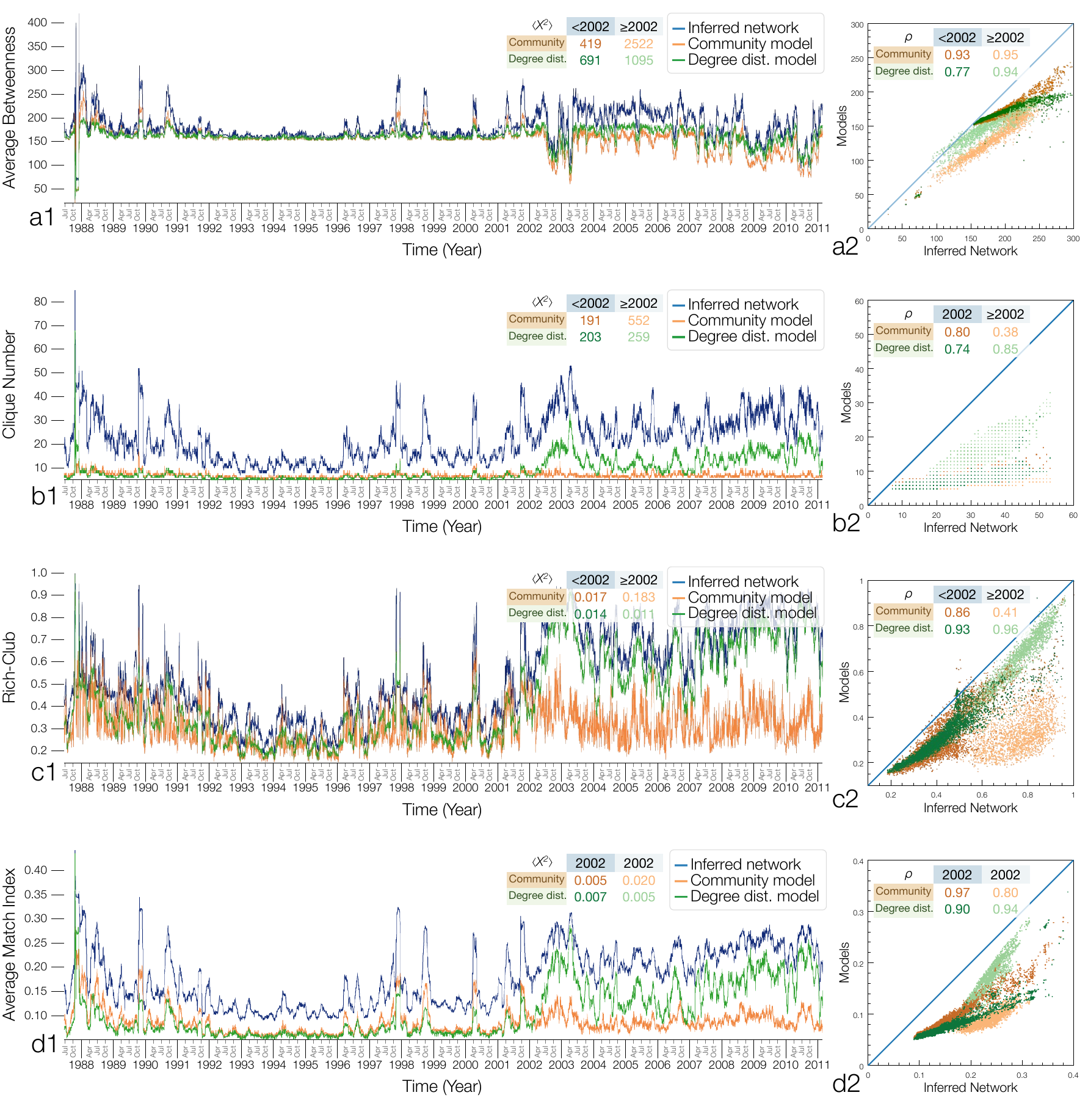} 
  \caption{Average betweenness, clique number, rich-club coefficient and average matching index obtained for the NYSE networks (blue), configuration null model (green) and community null model (orange). Plots on the left show the time evolution of the measurements. Plots on the right show the scatter plot between each measurement obtained from the inferred networks and the artificially generated networks. The blue line indicates a $y=x$ relationship. The average squared differences and Pearson correlation coefficients between the inferred and simulated time series are shown as tables included inside each plot. Both measurements are calculated separately for the $p_{1987\rightarrow2002}$ and $p_{2002\rightarrow2011}$ periods.}
  ~\label{fig:supplementaryMeasurementsPlots}
  \end{center}
\end{figure*}

\end{document}